\documentclass[twocolumn,prl,showpacs]{revtex4}
\usepackage{psfrag}
\usepackage{graphicx}
\usepackage{amsmath}
\usepackage{amssymb}

\begin{document}
\title{Forces and conductances in a single-molecule bipyridine junction}
\author{R. Stadler}
\affiliation{Center for Atomic-scale Materials Physics, \\
Department of Physics, Technical University of Denmark, DK - 2800 Kgs. Lyngby, Denmark}
\author{K. S. Thygesen}
\affiliation{Center for Atomic-scale Materials Physics, \\
Department of Physics, Technical University of Denmark, DK - 2800 Kgs. Lyngby, Denmark}
\author{K. W. Jacobsen}
\affiliation{Center for Atomic-scale Materials Physics, \\
Department of Physics, Technical University of Denmark, DK - 2800 Kgs. Lyngby, Denmark}

\date{\today}

\begin{abstract}
  Inspired by recent measurements of forces and conductances of
  bipyridine nano-junctions, we have performed density functional
  theory calculations of structure and electron transport in a
  bipyridine molecule attached between gold electrodes for seven
  different contact geometries.  The calculations show that both the
  bonding force and the conductance are sensitive to the surface
  structure, and that both properties are in good agreement with
  experiment for contact geometries characterized by intermediate coordination of the metal atoms
  corresponding to a stepped surface.  The conductance is mediated by the
  lowest unoccupied molecular orbital, which can be illustrated by a
  quantitative comparison with a one-level model.  Implications for
  the interpretation of the experimentally determined force and
  conductance distributions are discussed.
\end{abstract}
\maketitle

With the advance of new experimental techniques for the fabrication of
atomic-scale contacts, it has become possible in recent years to study
electron transport through a few or even a single molecule suspended
between metal electrodes.~\cite{joachim95,reed97,reichert02,smit02}
The atomistic details of the local contact geometry are crucial in
such experiments, as is evident from the low degree of reproducibility
of for example the measured I-V
curves~\cite{reichert02} or conductance traces~\cite{smit02}. Since the
exact contact geometry is difficult to control and characterize,
theoretical calculations are needed for interpreting and understanding
the results of the experiments.

Recently, the conductance of bipyridine molecules in a toluene
solution trapped between two gold electrodes has been measured
simultaneously to the force required to break the contact.~\cite{xu} 
The statistical data derived from
$\sim$1000 independent measurements show a quantization of both the
conductance and the rapture force, indicating that the contacts
consist of single molecules. Such bipyridine junctions are of
interest for two reasons: i) it is one of the few systems where fairly
reliable measurements exist of several properties (forces and conductances) for junctions of single aromatic
molecules, and ii) pyridine contacts represent an alternative to the
more commonly used thiol anchor groups~\cite{bilic}.  Conductances for
a bipyridine molecule between defect-free Au (111) surfaces have
recently been calculated by Hou et al.~\cite{hou} for a single
structure using a cluster
model for the description of the electrodes. However, in the
experiment the surface structures are not known, and here we take up
the issues of stability and the extent to which the force and conductance
distributions can be related to local structure.

We have performed plane wave-based density functional theory
(DFT)~\cite{dacapo} calculations for the conductance and breaking
force of a bipyridine-contact.  The contact between the surface and
the molecule is modeled by a variety of different structures based on
the (111) orientation of the Au fcc-lattice. By varying the
substrate/adsorbent distance rigidly we obtain total energy curves,
from which rapture forces can be derived for the different contact
geometries. A recently developed scheme for coherent electron
transport~\cite{transport} allows us to calculate the conductance of
the molecule within the same theoretical framework as employed for the
total energy calculations. 
Based on our data obtained for the different contact geometries we find that the force and 
conductance distributions observed in the experiment can only be understood in terms
of a high structure-selectivity for the contact geometry. 

\begin{figure}
\begin{center}
\includegraphics[width=1.0\linewidth,bb=0 0 784 502, clip]{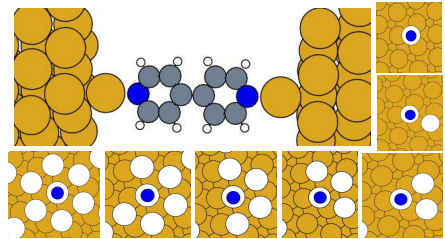}
\end{center}
\caption[system]{\label{fig1}A bipyridine molecule suspended between
  Au electrodes.  The insets illustrate the variation of the surface
  contact geometry, where clockwise from top-right to bottom-left a
  single ad-layer with one to seven Au ad-atoms (white
  spheres), where $N_{c}$ varies from three to nine (see text), on top of a flat Au (111) surface are shown.  Smaller dark
  spheres define the nitrogen positions.}
\end{figure}
Fig.~\ref{fig1} shows an example of the Au/bipyridine/Au
nano-junctions and defines the different electrode geometries
investigated in this article.  By using our DFT-plane wave
approach~\cite{dacapo} with the PW91 functional for exchange and
correlation~\cite{pw91} and an energy cutoff of 340 eV for the plane
wave expansion, we have established that bipyridine prefers to bond to
the Au (111) surface in an on-top position, and therefore all
calculations have been performed for an on-top contact. The supercells
consist of repeated slabs with seven Au layers in a (111) stacking
with 3x3 periodicity within the surface plane, where the slabs are
linked by one bipyridine molecule per unit cell.  As depicted in
Fig.~\ref{fig1} the number of Au atoms in the surface plane has been
varied from one to seven in such a way that the coordination number $N_{c}$ of the Au
atom directly connected to the molecule varies from three to
nine. The two structures, where the remaining vacancies in the surface plane are also filled for $N_{c}$=9, are not shown in Fig.~\ref{fig1}, since our calculations showed that occupying these atomic positions has no effect on the forces or conductances. The geometry of the isolated molecule has been
relaxed by total energy minimization, but the Au atoms have been fixed
to their positions in the bulk crystal structure throughout our study.
The torsion angle, $\alpha$, which arises if the two pyridine segments
within the molecule are rotated relative to each other has been
optimized and found to be around 25$^{\circ}$ when the molecules are adsorbed in the junction.  However, since the energy gain of this
rotation is only around 0.03 eV, we cannot predict the effect that a
solvent or finite temperatures might have on it.
We found that the influence of this rotation on the calculated
forces is negligible. For the conductance we have performed 
calculations for both $\alpha$=0$^{\circ}$ and 25$^{\circ}$.
An optimization of the Au-N bond length gave values between 2.12 and
2.42~\AA\ depending on the number of atoms in the surface plane but showed no dependency on $\alpha$. 

\begin{figure}
\begin{center}
\includegraphics[width=1.0\linewidth,bb=0 0 550 606,clip]{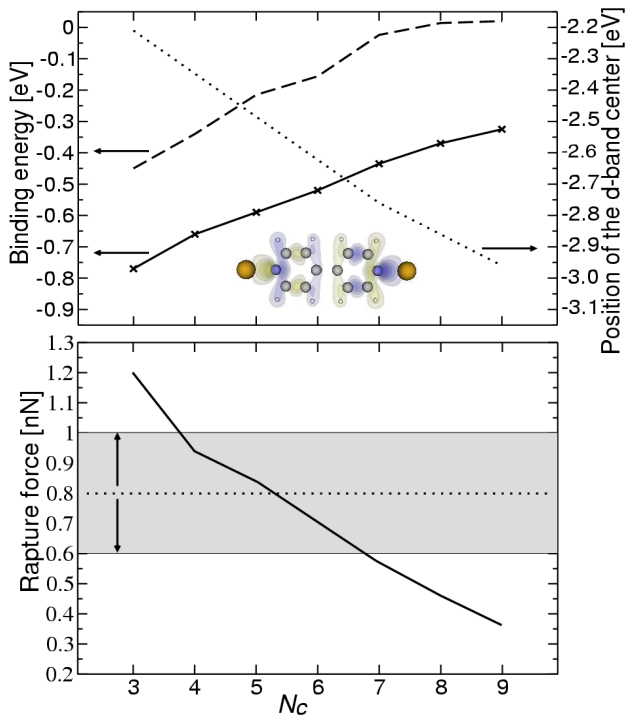}
\end{center}
\caption{\label{fig2}Top: Binding energies for a
  single N-Au bond as a function of $N_{c}$. The inset shows the HOMO.
  The solid (dashed) line marks calculations using the PW91
  [\protect\onlinecite{pw91}] (RPBE [\protect\onlinecite{RPBE}])
  XC-functional. The position of the {\it d}-band center with respect to the Fermi level is shown
  as a dotted line. Bottom: Rapture force as a function of $N_{c}$. The horizontal dotted
  line indicates the experimental value and the grey shaded area marks the region defined by the experimental error bars.}
\end{figure}

In the top panel of Fig.~\ref{fig2} the binding energies at the
optimal Au-N bond lengths for all the structures are shown. The
binding energies vary significantly with an almost linear dependence
on $N_{c}$, and their magnitude indicate intermediate-strength bonding
stronger than typical van der Waals bonding but weaker than covalent
bonding between open shell systems. Depending on the detailed choice
of exchange-correlation functional the binding energies may shift a
few tenths of an eV but the trend with metal coordination number is
unchanged (see Fig.~\ref{fig2}).  The same linear dependence of the
binding energy on $N_{c}$ has been found for the chemisorption of
oxygen and CO on Au chains and surfaces \cite{oxygen}, indicating that
the nature of the bonding is also covalent in the present case.
A closer
inspection of the surface states shows, that the trend in the binding
energies correlates with an upward shift of the $d$-states towards $E_F$ which is expected to enhance the energy gain of a
hybridization with the highest occupied molecular orbital (HOMO, see
inset) located at $\sim$ -4.6 eV.\cite{catalysis} These findings are in contrast with a recent
article, in which the adsorption of pyridine on Au (111) surfaces was
interpreted in terms of Van der Waals interactions~\cite{bilic} based
on more qualitative arguments such as the amount of charge buildup in
the center of the bonding region.

From the total energy curves obtained by varying the Au-N bond lengths, the
breaking forces can be calculated by taking the derivatives at the
inflexion point. As can be seen in Fig.~\ref{fig2} the
forces like the energies depend almost linearly on $N_c$. The rather large 
variation of the calculated breaking
forces should be compared with the experimental situation where a
quite sharp peak at a value of $0.8\pm0.2$ nN is found.~\cite{xu} This
distribution is consistent with our results if the
experimental situation results in the preference for a subset of
structures in the range $N_c$=4-7. Such a preference could be either thermodynamic or kinetic
in origin. Energetically bipyridine clearly favors to bond to metal
sites with low metal-coordination numbers according to the
calculations. On the other hand such metal defects may be
energetically costly to produce and therefore limited in number. Also
the experimental process in which the nanocontact is created may bring
the system out of local equilibrium and the resulting structures will
then depend on the evolution of the process.  Finallly it should be
noted that the calculations do not take potential effects of the
solvent into account.

The conductance of the bipyridine junctions is calculated using a
numerical method for phase-coherent electron transport which combines
a Green's function formalism with a plane wave-based DFT description
of the atomic and electronic structure of the entire system. The
Green's function of the central region, comprising the molecule and
part of the gold electrodes, is evaluated in terms of a basis
consisting of maximally localized Wannier functions which are defined
by an appropriate transformation of the Kohn-Sham eigenstates. 
Despite their optimal localization, the Wannier functions still have a 
fairly long-ranged but small tail, which for the systems studied in this article
we truncate at a distance of 19.4~\AA\ from the center of each function.
The couplings to the semi-infinite gold electrodes are included via
self-energy matrices which are also represented in the Wannier
function basis. A detailed discussion of the transport scheme can be
found in Ref.~\onlinecite{transport}. The central regions used in the
transport calculation are the same as the supercells applied in the
total energy calculations. These supercells contain $3\times 3$ atoms
in the directions perpendicular to the transport direction, and we
have found that $\bold k$-point sampling in the transverse plane is
important in order to obtain well converged results for the
conductance. Thus all calculations presented in the following have
been performed using a $4\times 4$ $\bold k$-point grid in the
transverse Brillouin zone. For a systematic study of the role of $\bold k$-point sampling for transport calculations see Ref.~\onlinecite{kpoints}.

\begin{figure}
\begin{center}
\includegraphics[width=1.0\linewidth,bb=0 0 550 706,clip]{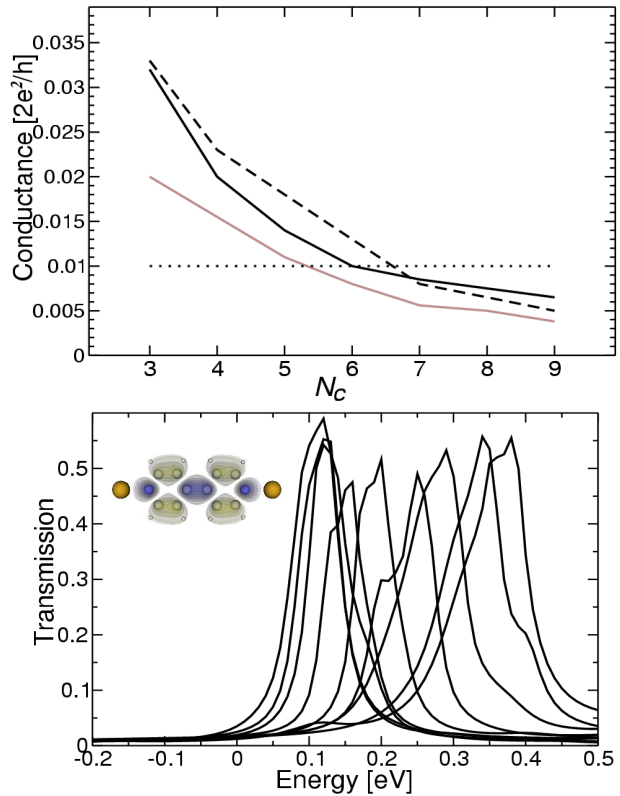}
\end{center}
\caption[channel]{\label{fig3}Top: Conductance in dependence on
  $N_{c}$ for torsion angles $\alpha$=0$^{\circ}$ (black line) and
  25$^{\circ}$ (grey line) in units of the conductance quantum $G_0=2e^2/h$. The dashed line refers to a simple model explained in the
  text. The horizontal dotted line shows the experimental value. 
  Bottom: Transmission functions for all nano-junctions
  ($\alpha$=0$^{\circ}$). The peaks are due to the LUMO, shown in the inset, 
  which rises continuously in energy as $N_{c}$ increases.}
\end{figure}

In Fig.~\ref{fig3} we have plotted the calculated transmission near
the Fermi level (which is taken as the zero point) for all geometries
outlined in Fig.~\ref{fig1}.  The top panel of Fig.~\ref{fig3} shows
the conductances for nano-junctions with $N_{c}$=3 to 9 for
$\alpha$=0$^{\circ}$ and $\alpha$=25$^{\circ}$ as black and grey solid
lines, respectively. 
The conductance is seen to vary significantly with coordination
number. However, the conductance 
for $\alpha$=25$^{\circ}$ is in good agreement with the experimentally observed value of
one percent of the conductance quantum $G_0=2e^2/h$.\cite{xu} It can be noted that this
is the same range of $N_c\sim$5, where also the forces in Fig.~\ref{fig2} match the measurements.
But as already
discussed in connection with the force distributions the actual
propensity of the different structures in the experiment is not known.

We now proceed to analyze why the conductance varies the way it does
with coordination number. The analysis is based on the calculations with
$\alpha=0^{\circ}$. 

Within our calculational scheme it is possible to define molecular
orbitals even when the molecule is coupled to the metal. The molecular
orbitals are constructed from Wannier functions located on the
molecule \cite{transport}. It is therefore also possible to study the
transport through a single or a subset of the molecular orbitals by simply removing the remaining orbitals from the basis set.
When we focus on the contributions coming from individual
molecular orbitals, we find that the single peak in the transmission
function around $E_F$ (bottom panel in Fig.~\ref{fig3}) is exclusively
due to the LUMO, whose wavefunction is shown in the inset. Since the
peak is quite narrow for all geometries, correlation effects, leading
to Coulomb blockade behaviour in other weakly bonded molecular
junctions~\cite{kubatkin}, could be relevant for its accurate
description. However, even in the "coherent limit" there is good agreement between 
our transport calculations and the experimental results.

One way of analysing the shape of the $T(N_{c})$ curve further is by
applying a simple one-level model~\cite{newns} with only one molecular
state (the LUMO) coupled symmetrically to the leads . Within this model the
transmission through a level at energy $E_0$ can be expressed as
$T(E)=\Delta^2/((E_0-E_F)^2+\Delta^2)$, where the peak-width
is defined by $\Delta=\pi V^2 \rho^0_g$. $V$ is the coupling
constant of the LUMO to its group-orbital,
\cite{transport,group} and $\rho^0_g$ is the density of states
of group orbital in the absence of coupling to the LUMO evaluated at $E_0$.  By taking the values
for $\rho^0_g, E_0$, and $V$ from the DFT calculation and using them to calculate $\Delta$, we
obtain the dashed line in Fig.~\ref{fig3}, which thus represents the
conductance when the electrons are allowed to travel between the leads via the LUMO only. 
The good agreement of this one-level model when
compared with our data further illustrates that the variation of the conductance is entirely controlled by the position of
the transmission peak due to the LUMO and that the effect of the interference with other orbitals as well as through-vacuum tunneling are negligible. 

But what is
governing the LUMO position? 
We investigated the variation of the energetic position of the lowest-lying 
molecular orbital (denoted MO1) with respect to the metal's Fermi level, in dependence on 
the distance between surface and molecule for all studied surface structures.
In this analysis it can be seen that at the bonding distance 
MO1 and the LUMO depend on $N_{c}$ in quantitatively the
same way. 
This pattern cannot be explained in terms of vacuum level alignment alone, since the variation
of the metal's workfunction with $N_{c}$ is parabolic with a minimum at $N_{c}$=5-6.
However, MO1 is $\sim$ 10 eV below the lowest-lying Au valence states,
and therefore the exchange or bonding interaction between the molecule and the surface do not
have a direct influence on the energy of this orbital. 
From that we can conclude that the dependence of MO1 as well
as the LUMO's energy on $N_{c}$ must be exclusively guided by rigid potential shifts, where 
we found that equilibrium charge transfer dominates at the optimal bonding distance and also Coulomb repulsion due
to the confinement of electrons at the interface plays a role. 
We will discuss these electrostatic effects and their structure-dependent influence on the aligment of molecular levels in detail in a separate publication.


In summary, we presented a theoretical analysis of the forces and
conducances of a bipyridine molecular junction. Since important
parameters of the experiments, such as the atomic configuration of the
electrode's surfaces are not known, we scanned a wide range of surface
structures and found that both the forces and conductances depend
crucially on their geometries, but our calculations agreed with the experiments in orders of
magnitude for both quantities. The narrow peaks
found in the statistics of the experiments, however, could not be directly
reproduced from our calculations, and the statistics of the measurements 
site-selectivity of the adsorption of the molecules has to play a major role. 
Our study suggests that varying surface structures might be an important ingredient for a
systematic understanding of the bonding and conductance properties of
molecular junctions.

The Center for Atomic-scale Materials Physics is sponsored by the
Danish National Research Foundation. We acknowledge support from the
Nano-Science Center at the University of Copenhagen and from the
Danish Center for Scientific Computing through Grant No. HDW-1101-05.


\end{document}